\documentstyle[12pt]{article}
\begin{document}

\newcommand{\be}{\begin{equation}}
\newcommand{\ee}{\end{equation}}
\newcommand{\txm}[1]{$#1$}
\newcommand{\az}{a_0^{zz\rightarrow zz}}
\newcommand{\lag}{{\cal L}}
\newcommand{\perpar}{\frac{g^2}{16\pi^2}\frac{m_H^2}{m_W^2}}
\newcommand{\ah}{\alpha_H}

\begin{titlepage}
\rightline{TURKU - FL - P9}

\vskip 3cm

\begin{center}
\bigskip

{\LARGE \bf Unitarity of the Standard Model at the higgs' resonance}\\
\bigskip
\bigskip

J. Sirkka\footnote{internet: sirkka@sara.cc.utu.fi}\\ Department of
Physics, University of Turku, SF-20500 Turku, Finland.\\
\bigskip
and\\
\bigskip
I. Vilja\footnote{internet: iiro@nordita.dk}\\
Nordita, Blegdamsvej 17, DK-2100 Copenhagen \O, Denmark\\
\bigskip\bigskip\bigskip

{\bf Abstract}
\end{center}
\begin{quote}
We compute a unitarity bound for higgs mass using one--loop corrected
s--wave partial amplitude for $Z_L Z_L \rightarrow Z_L Z_L$ scattering.
We use the equivalence theorem and show that the higgs mass has to be
less than $\approx 600$ GeV in order to save the (perturbative)
unitarity in the higgs' resonance region. We also discuss about the
validity of perturbation expansion in the symmetry breaking sector.
\end{quote}
\vfill
\end{titlepage}

The Higgs sector of the Standard Model is, unlike the other parts of it,
still poorly understood. The minimal version of the Standard Model has
one SU(2) Higgs doublet consisting of four real degrees of freedom. As a
result of  spontaneous symmetry breaking three of them become
longitudinal components $W_L^\pm$ and $Z_L$ of the massive weak gauge
bosons $W^\pm$ and $Z$, whereas one degree of freedom remains as a
physical particle $H$. The vacuum expectation value of this particle is
experimentally fixed, although the particle itself has not been
observed: the mass of the Higgs boson \txm{m_H}, or equivalently, the
value of the self--interaction coupling constant \txm{\lambda} of the
symmetry breaking sector is unknown.

By direct searches, the lower limit of the higgs mass has been able to
push as high as \txm{m_H>60} GeV \cite{lep}. The combined fit  of LEP
data and low energy experiments favor the higgs mass not to be more than
a couple of hundred GeV's \cite{fit} (assuming only the minimal Standard
Model) but the upper limit is not very conclusive. If \txm{m_H} is large
enough, perturbative calculations are not reliable any more because the
symmetry breaking sector becomes strongly interacting. The scattering
amplitudes of the would--be Goldstone bosons are related to the
scattering amplitudes of the longitudinal weak gauge bosons \txm{Z_L}
and \txm{W^{\pm}_L} according to so called equivalence theorem
\cite{CG,lqt}. It states that the scattering amplitudes of the
longitudinal gauge bosons are, in leading order of \txm{m_W/E}, where
\txm{E} is  the energy scale of the scattering, the same as for the
corresponding would--be Goldstone boson scattering amplitudes calculated
in the \txm{R_\xi}--gauge.

Our aim is therefore to study, how large \txm{m_H} may be in order to
consider perturbative calculations reliable.  An upper bound for the
higgs mass can be estimated by considering the limitations which the
requirement of (perturbative) unitarity of the scattering matrix puts on
its elements, in particular for the partial wave amplitudes. This
approach was first introduced by Lee, Quigg and Thacker \cite{lqt} who
found, using the tree level expression for the s--wave partial amplitude
$\az(s)$ and a rough unitarity bound $|\az(s)|< 1$, that \txm{m_H}
should  be less than 1 TeV in order to save the tree level unitarity in
the limit \txm{s\rightarrow\infty}. Later the high energy unitarity
bound was sharpened by Durand, Johnson and Lopez \cite{djl} (see also
\cite{others}) by including \txm{{\cal O}(g^2m^2_H/m^2_W)} one--loop
corrections  and using the more restrictive unitarity condition
\be
|\az(s)-\frac i2|<\frac 12\;. \label{ub}
\ee
Their result was that \txm{m_H} should be less than \txm{\simeq 400}
GeV in order to save the unitarity.

Our approach is to test the bound (\ref{ub}) near the resonance
\txm{s/m^2_H=1}. A good feature of this approach is that unlike the
bound in \cite{djl}, the bound we obtain to the higgs mass is
independent of the effective energy scale up to which the Standard Model
is considered to be a valid theory. It turns out that this approach
gives the unitarity bound for the higgs mass which is of the same order
of magnitude as the earlier bounds. Our aim is also to discuss, what is
consistent way to perform the perturbative expansion of the physical
quantities when the perturbation parameter \txm{g^2m^2_H/(16\pi^2m^2_W)}
is rather large.

We calculate the neutral Goldstone boson  \txm{zz\rightarrow zz}
scattering amplitude at one--loop level and  interpret the result as a
large \txm{m_H} approximation for the  \txm{Z_LZ_L\rightarrow Z_LZ_L}
amplitude according to the equivalence  theorem. In
\txm{Z_LZ_L\rightarrow Z_LZ_L} scattering \txm{s}, \txm{t} and \txm{u}
channels are all open while  in  e.g.\  \txm{w^+w^-\rightarrow w^+w^-}
scattering only \txm{s} and \txm{t} channels are open leading to a less
stringent unitarity bound, in general \cite{lqt,djl}. The interaction
Lagrangian reads

\be
\lag = -g_0^2\frac{m_{H_0}^2}{8m_{W_0}^2}\left( w^+w^-+\frac
12z^2+\frac 12H^2 +\frac{2m_{W_0}}{g_0} H+\frac{2m_{W_0}\delta
T}{g_0m_{H_0}^2}\right)^2,
\ee

where subscript 0 refers to bare quantities. The quantity \txm{\delta
T=\lambda_0v_0(v_0^2-\mu_0^2/\lambda_0)}, which vanishes at tree level,
cancels the tadpole term generated in the loop expansion. It can be
shown that \txm{\delta T} is related to the  Goldstone boson self-energy
at zero momentum transfer \txm{\Pi(0)}  by \cite{taylor} $\delta
T=-v_0\Pi(0)$. Therefore, our computational strategy is parallel to
\cite{djl,mw}:  we ignore all tadpole terms and subtract the zero
momentum Goldstone boson self-energy from all scalar self-energies,
which guarantees that the Goldstone bosons remain massless in the
presence of loop corrections, too. Furthermore, it is convenient to
perform all calculations in the Landau gauge where bare \txm{w^\pm} and
\txm{z} propagators have zero mass.

Our renormalization description is as follows: \txm{m_H} is taken to be
the physical higgs mass determined by the pole of the full propagator
(${\rm Im}\, \Pi \ll m_H^2$)
\be
m_H^2-m_{H0}^2-{\rm Re}\,\Pi(m_H^2)=0,
\ee
and the coupling constant is renormalized by defining
\be
\lambda=g^2\frac{m_H^2}{8m_W^2},
\ee
which is formally similar to the tree level relation. The parameters
\txm{g} and \txm{m_W} are renormalized so that only corrections
proportional to \txm{m_H^2/m_W^2} are taken into account in the
counterterms \txm{\delta g} and \txm{\delta m_W^2}. As a consequence of
this choice one obtains \txm{\delta g=0} and  \cite{marsir}
\be
\frac{\delta m_W^2}{m_W^2}=\frac 18 \perpar .
\ee
Furthermore, we have to specify a wave function renormalization for the
Goldstone bosons. We choose \txm{Z_z} to be defined as the residue
of the \txm{z}--propagator at its pole \txm{s=0}, whence
a calculation results
\be
Z_z=1-\frac 18 \perpar .
\ee

Near the resonance \txm{s/m_H^2=1} the amplitude changes very rapidly
so, that it is important to perform a full one loop calculation, in
particular to include various imaginary parts coming from three and four
point vertex functions. The complete one--loop s--wave partial amplitude
reads
\begin{eqnarray}
\az(s) & = & \frac 1{32\pi}\int_{-1}^1 d\cos\theta T(s,\cos\theta)
\nonumber \\
    & = & \frac{Z_z^2}{32\pi}\int_{-1}^1 d\cos\theta \Big [
          \Gamma_3(s)G(s)\Gamma_3(s)+\Gamma_3(t)G(t)\Gamma_3(t) \\
    & \ & \qquad +\ \Gamma_3(u)G(u)\Gamma_3(u) + \Gamma_4(s,t,u)\Big ]
          \nonumber,
\end{eqnarray}
where \txm{G} is the full, renormalized higgs propagator
\be
G(s) = \frac i{s-m_H^2-\left(\Pi_H(s)-{\rm Re}\,\Pi_H(m_H^2)\right)}
\ee

and \txm{\Gamma_3} and \txm{\Gamma_4} are  the proper three and four
point vertex functions, respectively.  The scattering angle \txm{\theta}
is related to the Mandelstam variables through the usual relations.
Introducing dimensionless integration and dynamical variables
\txm{w=-t/m_H^2} and \txm{x=s/m_H^2} the partial wave amplitude
\txm{\az} reads
\be
\az(x)=\frac{Z_z^2}{32\pi}\left( 2\Gamma_3^2(x)G(x)+\frac 4x\int_0^xdw
        \Gamma_3^2(-w)G(-w)+\frac 2x\int_0^xdw\Gamma_4(x,-w,w-x)\right).
\ee
To study the effects of the loop corrections we define the dimensionless
functions \txm{f_2}, \txm{f_3} and \txm{f_4} by
\begin{eqnarray}
\Gamma_3(x) & = & \frac{-ig}2\frac{m_H^2}{m_W}\left(1+\ah f_3(x)\right),
\nonumber \\
\Gamma_4(x,-w,w-x) & = & \frac{-3ig^2}4\frac{m_H^2}{m_W^2}
\left(1+\ah f_4(x,-w,w-x)\right), \\
G(x) & = & \frac i{m_H^2}\frac 1{x-x_0-\ah f_2(x)},\nonumber
\end{eqnarray}
where $\ah$ is the expansion parameter
\be
\ah=\perpar=0.42\, \left ({m_H\over \mbox{TeV}}\right )^2.
\ee
Note, that into the parameter \txm{x_0=1-\ah 3\pi i/8} are, actually,
all ${\cal O}(\ah)$ corrections to the pole of the propagator \txm{G}
included, because the function \txm{f_2} has the property
\txm{f_2(1)=0}.

Our results are summarized in Figure 1.
In the low energy region, \txm{0<x<0.5} where no resummation was made
for the \txm{s} channel propagator \txm{G(x)}, the partial wave
amplitude reads in first order in \txm{\ah}
\begin{eqnarray}
\az(x) & = &-\frac{g^2}{64\pi}\frac{m_H^2}{m_W^2} \Biggl [
          \left(3+\frac 1{x-1}-\frac 2x\ln(1+x)\right)\left(1-
          \frac 14\ah\right) \nonumber \\
    &  & + \ah\left(2f_3(x)+\frac{f_2(x)-\frac{3\pi i}8}{x-1}\right)
            \label{a0lowe} \\
    &  & + \frac{\ah}x  \int_0^xdw\left(3f_4(x,-w,x-w)-
          \frac{4f_3(-w)}{w+1}+\frac{2f_2(-w)-\frac{3\pi i}4}{(w+1)^2}
          \right)\Biggr].\nonumber
\end{eqnarray}
Although both functions \txm{f_3} and \txm{f_4} can be in general
evaluated only numerically, their asymptotic behaviour when their
arguments approach zero is rather easy to obtain. One can check that
logarithmic and constant terms cancel each other in accordance with the
low energy theorem \cite{lowthe}. In the low energy regime, however, we
can not interpret the Goldstone boson amplitudes as scattering
amplitudes of the longitudinal gauge bosons because the condition
$m_W/\sqrt s \ll 1$ of the equivalence theorem is not fulfilled.

In the resonance region, \txm{0.5<x<2}, the curve with dashed line in
Fig. 1 corresponds to \txm{\az(x)} evaluated in terms of a pole $x'_0$,
a residue $Z$ and a remnant part $R(x)$:
\be
\az(x)=\frac Z{x-x'_0}+R(x) .\label{laurent}
\ee
This form is adequate for the perturbation theory near a resonance where
usual perturbation expansion would lead to a divergence. Therefore, to
remove this singular behaviour, the leading $\ah$ correction of the pole
of the propagator has to been taken into account. The real part of the
self energy contributes, however, only to the ${\cal O}(\ah ^2)$
correction whereas the correction to the imaginary part is ${\cal
O}(\ah)$.  Thus in Eq.\ (\ref{laurent}), the terms \txm{Z}, \txm{x_0}
and \txm{R(x)} are evaluated in first order in \txm{\ah}, whence $x'_0 =
x_0$. The curve with solid line in Figure 1 corresponds to \txm{\az}
with proper \txm{s}--channel higgs propagator:
\begin{eqnarray}
\az(x) & = & -\frac{g^2}{64\pi}\frac{m_H^2}{m_W^2} \Biggl[
          \left(3-\frac 2x\ln(1+x)\right)\left(1-
          \frac 14\ah\right) \nonumber \\
    & \  & + \frac{1+2\ah f_3(x)-\frac 14\ah}{x-x_0-\ah f_2(x)}
\label{a0zfull}
           \\
    &  \ & + \frac{\ah}x\int_0^xdw\left(3f_4(x,-w,x-w)-
          \frac{4f_3(-w)}{w+1}+\frac{2f_2(-w)-\frac{3\pi i}4}{(w+1)^2}
          \right)\Biggr].\nonumber
\end{eqnarray}
The \txm{t} and \txm{u} channel propagators need not to be resummed,
because they are analytical in the \txm{-w<0} region. In practise, the
difference between these two forms is small in the resonance region,
even when the parameter $\ah$ is as large as
\txm{\ah=0.42\cdot(0.6)^2=0.15}. When \txm{x\gg 1}, situation is
completely different because of the logarithmic terms of the function
\txm{f_2}. For large $x$ they are important, which is analogous to the
concept of effective coupling constant which should be used as a
perturbation parameter. One expects that in high energy region one
should use Eq.\ (\ref{a0zfull}) instead of Eq.\ (\ref{laurent}).

Directly from the Fig.\ 1 we obtain the unitarity bound \txm{m_H<} 400
-- 600 GeV, though the unitarity is violated very softly. One also
concludes from Fig.\ 1 that the point \txm{x_c=s_c/m_H^2} where the
s--wave amplitude leaves the unitarity circle is a decreasing function
of the higgs mass. For \txm{m_H=400} GeV \txm{x_c=1.7}, while for
\txm{m_H=600} GeV \txm{x_c=1.4}. In the case of high energy unitarity
limit one would interpret \txm{\sqrt{s_c}} to be the effective scale up
to which the Standard Model is valid effective theory and above which
new physics is expected to show up to save the perturbative unitarity.
However, if the unitarity limit is near the higgs' resonance,
\txm{\sqrt{s_c}\approx m_H}, this interpretation is not possible
\cite{dasneu}. The breaking of the unitarity is a sign of strong
interaction physics and one expects that if the breaking scale
\txm{\sqrt{s_c}} is not too far from \txm{m_H}, the Higgs scalar is not
a propagating state by itself but form a bound state opening a new
channel. It pulls the s--wave amplitude inside the unitarity circle.

One should notice, that \txm{{\cal O}(m_W/\sqrt{s})} --correction
implied due the use of the equivalence theorem change the slope of the
curves in Figures 1 by 20 \% at most. One expects therefore that this
correction does not chance much the point where the partial wave
amplitude curve leaves the unitarity circle, if the curve crosses the
circle perpendicularly enough. To be on the safe side, the unitarity
bound near the higgs' resonance should be taken \txm{m_H \leq 600} GeV.
Although the first order corrections lower the unitarity bound from that
calculated in the tree approximation, the second order corrections are
not expected to change first order results. When \txm{m_H=600} GeV, the
perturbation parameter \txm{\ah} has the value \txm{\ah=0.15} which
means that near the higgs' resonance they are comparable to the
corrections implied by the equivalence theorem. Furthermore, inclusion
of the fermions, the \txm{t}--quark in practice, affects the unitarity
bound near the resonance ($m_H = 600$\ GeV) about 20 \%, too. The decay
width of the higgs, in other words the imaginary part of the pole
parameter \txm{x_0}, is expected to be the most sensitive to
\txm{m_W/m_H} corrections implied by the equivalence theorem \cite{mw}
and top--quark effects. The inclusion of them results \txm{x_c=1.8} for
\txm{m_H=400} GeV and \txm{x_c=1.5} for  \txm{m_H=600} GeV. One expects
that these corrections to the parameters \txm{Z} and \txm{R(x)} in
addition to \txm{x_0} tend to rise the value of \txm{s_c} slightly but
the unitarity bound for the higgs mass remains essentially unchanged.
Thus we may conclude that the  higgs mass should not exceed 600 GeV in
order to save perturbative calculations in the Higgs sector of the
Standard Model.

\newpage
\section*{Figure caption}

{\bf Figure 1.} The behaviour of the s-wave amplitude \txm{\az} for
\txm{0<x<2}. {\bf (a)} \txm{m_H=300} GeV, {\bf (b)} \txm{m_H=400} GeV
and {\bf (c)} \txm{m_H=600} GeV. The tic--marks denote the values \txm{x
= 0.5,\ 1.0} and 1.5. The curve with dashed line is
Laurent--approximated s-wave amplitude (\ref{laurent}) and that with
solid line is the one from (\ref{a0zfull}). The low energy
part of the curve, \txm{0<x<0.5}, is calculated from (\ref{a0lowe}).


\newpage

 \begin{thebibliography}{10}
\bibitem{lep} L.\ Rolandi, Talk presented in XXVI International
Conference on High Energy Physics 1992, Dallas, Texas.
\bibitem{fit} J.\ Ellis, G.\ L.\ Fogli and E.\ Lisi, Phys.\ Lett.\ {
\bf B274} (1992) 456.
\bibitem{CG}  J.\ M.\ Cornwall, D.\ N.\ Levin and G.\ Tiktopoulos, Phys.\ Rev.\
              {\bf D10} (1974) 1145, M.\ S.\ Chanowitz and Mary K. Gaillard,
              Nucl.\ Phys.\ {\bf B261} (1985) 379.
\bibitem{lqt} B.\ W.\ Lee, C.\ Quigg and H.\ B.\ Thacker, Phys.\ Rev.\ {\bf
D16}
              (1977) 1519.
\bibitem{djl} L.\ Durand, J.\ M.\ Johnson and J.\ L.\ Lopez, Phys.\ Rev.\
Lett.\
              {\bf 64} (1990) 1215; Phys.\ Rev.\ {\bf D45} (1992) 3112.
\bibitem{others} G.\ Passarino, Nucl.\ Phys.\ {\bf B343} (1990) 31; L.\ Durand,
              J.\ M.\ Johnson and P.\ N.\ Maher, Phys.\ Rev.\ {\bf D44} (1991)
              127.
\bibitem{taylor} J.\ C.\ Taylor, {\it Gauge Theories of weak Interactions}
              (Cambridge University Press, Cambridge, 1976).
\bibitem{mw}  W.\ J.\ Marciano and S.\ S.\ D.\ Willenbrock, Phys.\ Rev.\ {\bf
              D37} (1988) 2509.
\bibitem{marsir} W.\ Marciano and A.\ Sirlin, Phys.\ Rev.\ {\bf D22} (1980)
              2695.
\bibitem{lowthe} M.\ Chanowitz, M.\ Golden and H.\ Georgi, Phys. Rev.\ Lett.\
              {\bf 57} (1986) 2344; Phys.\ Rev.\ {\bf D36} (1987) 1490.
\bibitem{dasneu} R. W. Dashen and H. Neuberger, Phys. Rev.\ Lett.\ {\bf 50}
               (1983) 1897.
\end{thebibliography}
\end{document}